# Submerged Ancient Indian continent in the Bay of Bengal-inference from ambient noise and earthquake tomography

Gokul Kumar Saha[1*], S. S. Rai, K. S. Prakasam and V. K. Gaur

[*]Corresponding author: Gokul Kumar Saha (Email: kumarsahagokul123@gmail.com)

Affiliation

[1]Department of Earth and Climate Science, Indian Institute of Science Education and Research, Pune, Dr. Homi Bhabha Road, Pashan, Pune 411008, India

**Abstract**

We present evidence for an uninterrupted continuation of Indian continental lithospheric mantle into the adjoining Bay of Bengal to a distance of 400-500 km away from the passive margin. The inference is based on the shear wave velocity image of the uppermost mantle beneath the Bay of Bengal, Bangladesh, and the adjoining Indian craton, computed using ambient noise and earthquake waveform data. The Indian lithospheric mantle is characterized by a shear wave velocity of ~ 4.1-4.3 km at the Moho depth of 35-40 km, progressively increasing to ~4.5-4.7 km/s at least up to a depth of 140 km. This velocity structure continues uninterrupted to about $86^{\circ}$ E in the Bay of Bengal. Further east, the thickness of the lithospheric lid decreases to ~90 km and is underlain by reduced shear wave velocity (~4.1-4.3 km/s) in the uppermost mantle. We postulate that the Indian craton is embedded in the western Bay of Bengal and the continent-ocean boundary lay around $86^{\circ}$ E. The craton possibly submerged soon after the India-Australia-Antractica rifting at around 136 Ma. The significantly reduced shear wave velocity beneath the eastern Bay of Bengal may be due to reheating of the mantle as a consequence of its interaction with the Kergulean hotspot around 90 Ma.

**Introduction**

The Bay of Bengal (Figure 1) is generally agreed to have evolved due to the breakup of India from Antarctica and subsequent seafloor spreading (Norton and Sclater, 1979). The initial India-Australia-Antarctica rifting created the western Bay of Bengal around 136 Ma followed by the evolution of the eastern Bay of Bengal due to northward ridge jump at around 118 Ma (Gaina et al., 2007). The boundary between the two segments is proposed to be around $85^{\circ}$-$86^{\circ}$ E (Talwani et al., 2016). The other important features in the Bay of Bengal include the linear $85^{\circ}$ E and $90^{\circ}$ E ridges (Figure 1) widely considered as the traces of the Crozet and Kerguelen hot spots (Kent et al., 1992). To the north of Bay of Bengal is the Bangladesh formed over the oceanic crust and currently covered with large sedimentary thickness transported by the Ganges and Brahmaputra rivers (Johnson and Nur-Alam, 1991). The sediment column thickness increases to ~22 km beneath the Bangladesh shelf that progressively reduces to 3 to 4 km in the southern Bay of Bengal (Curray, 1994). Between Bangladesh and the Indian craton lies the Bengal basin that has developed largely over a remnant ocean basin. It is proposed that the Bay of Bengal has 5 to 6 km thick oceanic crust south of $15^{\circ}$ N (Curray et al., 1982).

Scientific opinion remains divided on the nature of the crust beneath the Bay of Bengal. The controversy initially emerged from the velocity-depth modeling of surface wave group velocity data of Brune and Singh (1986) who argued for a continent like thick crust beneath the Bay of Bengal. Their inferred Moho depth in the northern part of the basin is 5 to 10 km deeper than the earlier studies (Curray et al., 1982). To resolve this discrepancy, Brune et al. (1992) reanalyzed their surface wave data with constraints from high-frequency Sn wave propagation along several paths across the Bay of Bengal. Their reinterpretation suggests a 22 km thick sedimentary basin under the northern Bay of Bengal. The lower 6 km layer of the sedimentary basin is characterized by the P-wave velocity of ~6.5 km/s. This increased velocity is proposed to be due to the high-pressure metamorphism. This layer was earlier misidentified by Brune and Singh (1986) as oceanic crust leading to the interpretation of an unusually thick crust. The same argument was used by Mitra et al. (2011) to infer the oceanic nature of the crust beneath the Bay of Bengal from their velocity tomography results. The proposition of oceanic crust in the northern Bay of Bengal has been recently questioned by Sibuet et al. (2016) and Rangin and Sibuet (2017) based on several multi-channel seismic data acquired in the region. These studies suggest the presence of a thin (15 km thick) continental crust injected by Mesozoic volcanism, a view that is contradicted by Talwani et al. (2016).

Contra these studies, our knowledge of the state of sub-crustal lithosphere mantle beneath the Bay of Bengal is limited. Lithosphere, representing the plate, is characterized by high-velocity lid overlying the low-velocity zone corresponding to the asthenosphere (Eaton, 2009). The concept of the lid and the low-velocity zone have evolved in the past from different processes like partial melting (Gutenberg, 1959; Kawakatsu et al., 2009), mineral physics considerations (Anderson and Sammis, 1969), chemical changes (Regan and Anderson, 1984), and the presence of hydrated phases (Karato, 2012). The lithosphere-asthenosphere boundary (LAB) is classically associated with the depth of the $1300°C$ isotherm (Artemieva, 2006) and can be related to the $V_{SV}$ parameter at the top of the low-velocity zone. The velocity drop at the LAB varies from 5% to 10% depending on the age of the ocean. The effect of increasing lithospheric age is to shift the low-velocity zone to deeper depths and increase the value of minimum velocity. Considering an ambient shear wave velocity of 4.7 km/s, for a 100 Ma oceanic lithosphere the minimum shear wave velocity is expected to be around 4.3 km/s (Stixrude and Lithgow-Bertelloni, 2005). The thickness of most of the oceanic lithosphere increases with the square root of its age-about 80 km for an ocean at an age of 100Ma (Doin and Fleitout, 1996). In the northwestern Pacific ocean with seafloor age of 150-160 Ma, Nishimura and Forsyth (1989) inferred the lithospheric thickness of about 90 km using Rayleigh wave phase velocity data. Their results indicate high shear wave velocity of 4.7 km/s ± 0.07 km/s in the upper lithosphere that extends to ~65 km depth, followed by a negative velocity gradient with a minimum of $4.30 \pm 0.09$ km/s at 150 km depth. For 100-130 Ma old oceans, Kawakatsu et al. (2009) and Kumar and Kawakatsu (2011) suggest LAB at ~70 km depth.

We briefly discuss our current state of knowledge regarding the lithosphere beneath the Bay of Bengal. Based on the efficient propagation of Sn wave along a few paths, Brune et al. (1992) argued for cold uppermost mantle beneath the Bay of Bengal. Bhattacharya et al. (2013), inverted five inter-station surface wave phase velocity data to map the path-averaged LAB at a depth of about 110 km. Similar thickness of high velocity mantle is inferred from the global tomographic map computed using group velocity dispersion data (Shapiro et al., 2008). A combined modelling of gravity and geoid data suggest thickness of ~120 km in the northern Bay of Bengal (Rao et al., 2016). These studies are, however, either limited in extent or averaged over the region. Lateral variability of the structure of the Bay of Bengal lithosphere, therefore, remains speculative.

In this research paper, we present a detailed 3-D shear wave velocity image of the lithospheric mantle to a depth of 140 km beneath the Bay of Bengal and Bengal basin/Bangladesh and compare them with the adjacent Indian craton. The velocity image is used to understand the seismic character of the lithospheric mantle beneath the Bay of Bengal and to place a constraint on the continent/oceanic lithosphere dichotomy. We use here surface wave tomography approach combining ambient noise and earthquake waveform to generate the velocity image. The method and its application in different geological environments has been reviewed in Thurber and Ritsema (2015) and Campillo and Roux (2015).

**Data and Methodology**

We use Rayleigh wave group velocity data retrieved primarily from cross-correlation of ambient noise recorded over 683 seismic stations covering south Asia and adjoining north Indian ocean along with data from China and neighbouring regions (Figure 2a). To improve lateral resolution we have supplemented the ambient noise data with earthquake surface waveform data from 417 earthquakes (M>5.5, depth 10-95 km) between $10^{°}$S to $50^{°}$N and $31^{°}$E to $130^{°}$E, recorded by 209 seismic stations over India (Figure 2b).

The data processing of ambient noise to extract Rayleigh wave group velocity is accomplished following the techniques of Bensen et al. (2007). All the vertical component waveforms data recorded on 683 seismographs are decimated to one sample/s and cross-correlated among the stations operated during the same time interval. The cross-correlated functions (CCF) are averaged to generate a symmetric signal, which is further stacked for the entire duration of operation to increase the signal to noise ratio (SNR). In subsequent analysis, we use the waveforms with S/N>15 and the source-receiver distance more than three times the wavelength (Bensen et al., 2007). We compute the fundamental mode Rayleigh wave group velocity from CCFs and also the earthquakes waveforms recorded over seismic stations using the multiple filter taper analysis approach (Herrmann and Ammon, 2004). The combined ray coverage using ambient noise and earthquake data for every $1^{°} \times 1^{°}$ grid is presented in Figure S1 for selected time periods. The Rayleigh wave travel times along multiple paths are then converted into group velocity maps at different time periods following Barmine et al. (2001).

The horizontal resolution of the tomographic maps is investigated using the checkerboard resolution test in which the input synthetic model comprises an alternating pattern of higher and lower wave velocity (Rawlinson and Spakman, 2016). Using the observed ray-path geometry, we compute the synthetic group velocity travel time at different time periods and invert them to reconstruct the input velocity map. Here we use a discrete spike test involving a sparse distribution of spikes. We used square cells of size $0.5^\circ$, $1^\circ$, and $2^\circ$ with alternate ±6% velocity perturbation. The test is performed for Rayleigh wave group velocity at selected time periods from 10s to 70s. The minimum cell size of $1^\circ$ could be resolved for the study region. Figure S2 shows the recovered velocity for every $1^\circ \times 1^\circ$ horizontal grid.

**Surface wave dispersion maps**

Having established the reliability of velocity imaging methodology, we present surface wave velocity maps (Figure 3) at time periods from 10s to 70s over the Bay of Bengal, Bengal basin/Bangladesh and the adjoining Indian craton. Since the group velocity inversion at any period is performed from independent travel time data sets, continuity of features in imaged velocity maps at subsequent time periods suggests that the anomalies accurately represent the structural features. To establish a correspondence between the group velocity at any time-period and depth of investigation in the Earth, we present the group velocity sensitivity kernel for different time periods assuming a layered Earth model (Figure S3).

The group velocity map at the time-period of 10s-15s (Figure 3) is an average for the velocity in the approximate depth range of 4-15 km with a peak sensitivity at a depth of ~8-12 km. In this depth range, we observe strong variation in the morphotectonic features of the Earth differentiating crystalline upper crust, sedimentary basins, fold belts, oceanic crust, etc.

(Pasyanos, 2005). The Indian craton, at the time period of 10s-15s, has a group velocity of ~3.0-3.1 km/s representing crystalline crust. Progressively, the velocity reduces near the craton boundary to ~2.8-2.9 km/s in the Bengal basin/Bangladesh and the Bay of Bengal. Bangladesh is characterized by progressive lowering of velocity from the west to the east which could be a consequence of variation in sediment thickness of 3-5 km in the west to 10-12 km in the east (Lindsay et al., 1991).

The group velocity at the 20s period has an average group velocity of 3.1 km/s in the continental region, decreasing progressively to 2.8 km/s beneath the Bengal basin/Bangladesh. The Bay of Bengal has a group velocity of 3.2 to 3.4 km/s. The depth sensitivity for 20s period Rayleigh wave is between 10 km and 35 km corresponding to uppermost mantle beneath the ocean in contrast with lower crust in the continent. Usually, the 20s period group velocity response shows strong discrimination between the continent and the ocean because of the difference in their crustal thickness (~35 km vs ~10 km). Presence of similar group velocity in Indian continent and the adjoining Bay of Bengal suggests a thicker crust beneath the oceanic region.

At a higher time-period of the 30s, the Indian continent and the Bay of Bengal have distinct group velocity of 3.2-3.4 and 3.6-3.8 km/s respectively. The response at the 30s corresponds to a velocity structure at a depth of 15-50 km with the maximum resolution at 30-35 km depth. It is possible that due to the presence of thicker crust beneath the Bay of Bengal, the distinction between the continent and the ocean is delayed to the 30s group velocity data. With a further increase in the time-period to 50s, corresponding to the velocity response in the depth of 40-100 km, the average group velocity varies between 4.0 and 4.2 km/s. These features continue even at the deeper time-period of the 70s. A noteworthy observation is the presence of a high velocity (>4.2 km/s) region covering most of the adjacent Indian craton (Singhbhum, Bastar and eastern Dharwar) representing their cold upper mantle.

**3-D shear wave velocity structure**

To quantify the depth variation of shear wave velocity, we extract the Rayleigh wave group velocity dispersion data from 10s to 70s for each grid node of $1° × 1°$ and invert them for shear wave velocity as a function of depth using the linearized least squares inversion (Herrmann and Ammon, 2004). The starting model for the velocity inversion consists of a stack of isotropic layers with constant shear wave velocity of 4.5 km/s. The layer thickness is one km up to 10 km depth, followed by two km thick layers from 10 to 50 km depth, and five km thick layer extending from 50 to 140 km depth. During the inversion, constant velocity and density are assumed in each layer. The density is calculated using the relation $\rho = 0.32V_p + 0.77$ (Nafe and Drake, 1963). We have performed inversion with other initial velocity models like AK135 and PREM and found an insignificant change in the result. In case of data from the oceanic region appropriate thickness of water column was considered. We repeat the inversion 20 times to compute the final shear wave velocity-depth model at each node. The 1-D velocity-depth is generated at every $0.5°$ interval and stitched to get a 2-D velocity-depth section and velocity maps at different depths.

To study the variation of velocity from the continent to the ocean, we plot the shear wave velocity-depth section along the longitude at different latitudes from $12°$ to $25°$ N (Figure 4A to G). Beneath the Indian craton, we observe high velocity (upto 4.7 km/s) continuing to a depth of 140 km with no sign of velocity reversal. This signature continues laterally beneath the Bay of Bengal until $86°$ E beyond which, we observe velocity reversal with depth with minimum at a depth of about 90 km representing LAB. Beneath Bangladesh, the velocity reversal is observed at a depth of ~110 km. In summary, the velocity-depth sections and maps suggest two distinct patterns: a layered lithosphere mantle similar to the Indian craton beneath the region west of 86$^{\circ}$ E (western Bay of Bengal) and a thin (<90 km) lithospheric mantle to the east of 86$^{\circ}$ E in the region widely known as eastern Bay of Bengal. The underlying asthenosphere beneath the Bengal basin and Bangladesh (Figure 4 F and G) has marginally higher shear wave velocity (4.3-4.4 km/s) than that beneath the eastern Bay of Bengal (Vs ~4.1-4.3 km/s) (Figure 4 A to E). This feature is well reflected in velocity map of 80, 100 and 120 km depth (Figure S4).

**Geological interpretation of velocity-depth sections**

Theoretical considerations and other observations suggest a lithospheric thickness of about 90-100 km for oceans of age over 90 Ma. Beneath the Indian craton, we observed a two-layer lithospheric mantle with a maximum shear wave velocity of 4.7 km/s, extending at least up to 140 km beyond which we do not have depth resolution. With these constraints, we examine the boundary of the Indian continent vs oceanic lithosphere. We observe the signature of Indian continental lithosphere continuing beyond the shoreline and uninterrupted traceable up to ~86$^{\circ}$ E beneath the Bay of Bengal. This seismically imaged thick and old lithosphere beyond the coastline up to over 400-500 km into the Bay of Bengal suggests that the northern Indian ocean is a mix of continent and ocean. Similar high-velocity thick lithosphere beneath the ocean basins has been mapped elsewhere (King and Ritsema, 2000; Deen et al., 2006; Begg et al., 2009; Kaban et al., 2016). The phenomena responsible for the presence of thick continental cratonic lithosphere found beneath the oceans remain unresolved. It is argued that either the cratonic mantle lithosphere is not permanently attached to the drifting continental lithosphere and could detach during the process of continental movement (Wang et al., 2017) or it has shifted due to basal drag induced by mantle flow (Kaban et al., 2016).

Along most of the profiles, we observe more than 140 km thick high-velocity lid continuing until about $86^{\circ}$ E and then progressively decreasing to 75-90 km further east. This is in contrast to a mono-phase rifting model of McKenzie (1978) where the breakup is instantaneously resulting in the juxtaposition of continental and oceanic crust. The contact between these two types of crusts is often assumed to be sharp and marked by a magnetic anomaly. This classical model is significantly revised to show that breakup is gradational rather than a sharp event as evident along the Iberia-Newfoundland conjugate margins (Pe´ron-Pinvidic et al., 2007). A more detailed analysis by Huismans and Beaumont (2011) suggested that Iberia-Newfoundland type observations could be explained by depth dependent extension in which crust and lithosphere are firmly bonded. Here, the crust breaks first while the lithospheric mantle is still necking and progressively thins with time. This would lead to progressive thinning of continental lithosphere away from the continent, a signature also observed east of $86^{\circ}$ E in the Bay of Bengal.

Another interesting feature of the Bay of Bengal emerges from an examination of the distribution of minimum Vs in the depth section. We observe that the western Bay of Bengal has velocity signature of a stratified lithosphere akin to the adjoining Indian Archean cratons. In contrast, the eastern Bay of Bengal has a significantly lower Vs of 4.1 to 4.3 km/s in the depth of 90-120 km. The contact between the western and the eastern Bay of Bengal happens to be around $86^{\circ}$ E. Talwani et al. (2016) proposed that the eastern basin is younger than 118 Ma. They, however, could not place lower bound on the age of the eastern ocean basin. Beneath the Bengal basin/Bangladesh, the sub-lithospheric low-velocity is 4.3-4.4 km/s while it is 4.1-4.3 km/s in the eastern Bay of Bengal. As discussed above, the effect of decreasing lithospheric age is to decrease the value of minimum velocity. We present the minimum shear wave velocity over the Bay of Bengal and adjoining region in Figure 5. The minimum velocity map shows very distinct low velocity 4.1-4.3 km/s of the eastern Bay of Bengal in contrast with 4.4-4.6 km/s over the western Bay of Bengal continuing into the Indian craton. The local minimum velocity is related to the age of the lithosphere by $V_{smin} = V_{s0} - \frac{1}{\left(1.63 + 0.16t^{\frac{1}{2}}\right)}$, where $V_{s0} = 4.77$ km/s and $t$ is in Ma (Stixrude and Lithgow-Bertelloni, 2005). For a 100 Ma ocean, the Vs minimum is about 4.3 km/s.

The reduced Vs of 4.1 km/s beneath the eastern Bay of Bengal (Figure S4, depth 100, 120 km map and Figure 5) suggests significant high viscosity and hence, a higher degree of melting in the asthenosphere. Presently we do not have a clear idea of what led to such heating of the basin. It may be noted that in the northernmost part of Bangladesh, lies the Rajmahal and Sylhet traps that possibly originated from a Kerguelen hotspot (Curray and Munasinghe, 1991; Kent et al., 2002). As India moved northward, the trail of this hotspot is seen as 90° E ridge. Interaction of this hotspot could have led to reheating of the eastern basin.

**Conclusions**

This is the first study to resolve the presence of two distinct lithospheric features in the Bay of Bengal: the western basin (west of 86° E), a thick and high velocity layered lithosphere (> 140 km, Vs ~4.5-4.7 km/s) similar to the Archean Indian craton; and a thinner lithosphere (~90 km) underlain by a low-velocity layer (Vs~4.1-4.3 km/s) possibly representing asthenosphere to the east of 86° E representing the eastern Bay of Bengal. Given the uninterrupted continuation of Indian lithosphere velocity signature up to 86° E, we speculate the possibility of continuation of the Archean craton beneath the western Bay of Bengal and the continent-ocean boundary at 86$^\circ$ E. The western Bay of Bengal possibly represent the submerged Indian craton. This could have happened after the India-Antarctica-Australia rifting at 136 Ma. The eastern Bay of Bengal was subjected to multi-phase depth dependent rifting which also evidenced in its distinct lithospheric structure. The reliability of the hypothesis proposed here could be tested through petrological analysis of mantle rock samples and improved seismic imaging using broadband ocean bottom seismographs.

**Data and Resources**

Indian seismic waveform data were acquired from the National Geophysical Research Institute, Hyderabad and Fourth Paradigm Institute, Bangalore. We are grateful to the staff of above institutes for their help in data acquisition. Field deployment at the NGRI was made under several research programs with S. S. Rai as the principle investigator. Seismic waveform data from other stations were obtained from the IRIS Data Management Center. The data used here is available on request from one of the authors (S. S. Rai).

**Acknowledgments**

This research was supported by the Indian Ministry of Earth Sciences (project-MoES/P.O.(Geo)/58/2016/Dr. S. S. Rai) and the Indian Department of Science and Technology (JC Bose National fellowship to Rai). We sincerely thank Bob Hermann (Saint Louis University, MO), Mike Ritzwoller (CU Boulder), Mikhail Barmine (CU Boulder), and K. S. Krishna (University of Hyderabad, India) for their advice during various stages of this research. We thank Mikhail Barmine for providing computer code for velocity imaging.

**Mailing address for each author**

1. Gokul Kumar Saha

Department of Earth and Climate Science,

Indian Institute of Science Education and Research, Pune,

Dr. Homi Bhabha Road, Pashan, Pune 411008, India

Email: kumarsahagokul123@gmail.com

Mobile No.: 9849954820

2. S. S. Rai

Department of Earth and Climate Science,

Indian Institute of Science Education and Research, Pune,

Dr. Homi Bhabha Road, Pashan, Pune 411008, India

Email: shyamsrai@gmail.com

Mobile No.: 9890322705

3. K. S. Prakasam

#2-19-22/2, Street no. 4, Kalyanapuri, Uppal, Hyderabad 500039, India

Email: ksprakasam@gmail.com

Mobile No.: 9440515319

4. V. K. Gaur

CSIR-Fourth Paradigm Institute, NAL Belur Campus, Bangalore-560037, India

Email : gaurvinod36@gmail.com

Mobile No.:9453629839

**List of Figure Captions**

**Figure 1:** Geological map of the study region depicting individual terrains of the region superimposed on the free air gravity anomaly.

**Figure 2:** (a) Map shows the distribution of 683 seismic stations from India and the neighboring countries used to compute cross-correlation function from the vertical component ambient noise recording. This contributed to more than 21600 ray paths. (b) Additional group velocity data were generated using 417 earthquakes (red star) and the corresponding 209 seismic stations (black reverse triangles).

**Figure 3:** Fundamental mode Rayleigh wave group velocity maps at periods of the 10s, 15s, 20s, 30s, 50s, and 70s. Group velocity is contoured at an interval of 0.2 km/s.

**Figure 4:** Shear wave velocity-depth section along west to east from Indian craton to the Bay of Bengal/Bangladesh. Velocity sections are generated at different latitudes from 12$^\circ$ N to 25$^\circ$ N (marked as profile from A to G). Shear wave velocity is contoured at an interval of 0.2 km/s. Major geological domains and tectonic boundaries are marked on top of individual figures.

**Figure 5:** Map view of minimum shear wave velocity in the upper mantle beneath Bay of Bengal and adjoining Indian craton. Note the continuity of high velocity mantle beneath the Indian craton into the western Bay of Bengal. Interpreted Indian continent-ocean lithosphere boundary is shown by a thick black line.

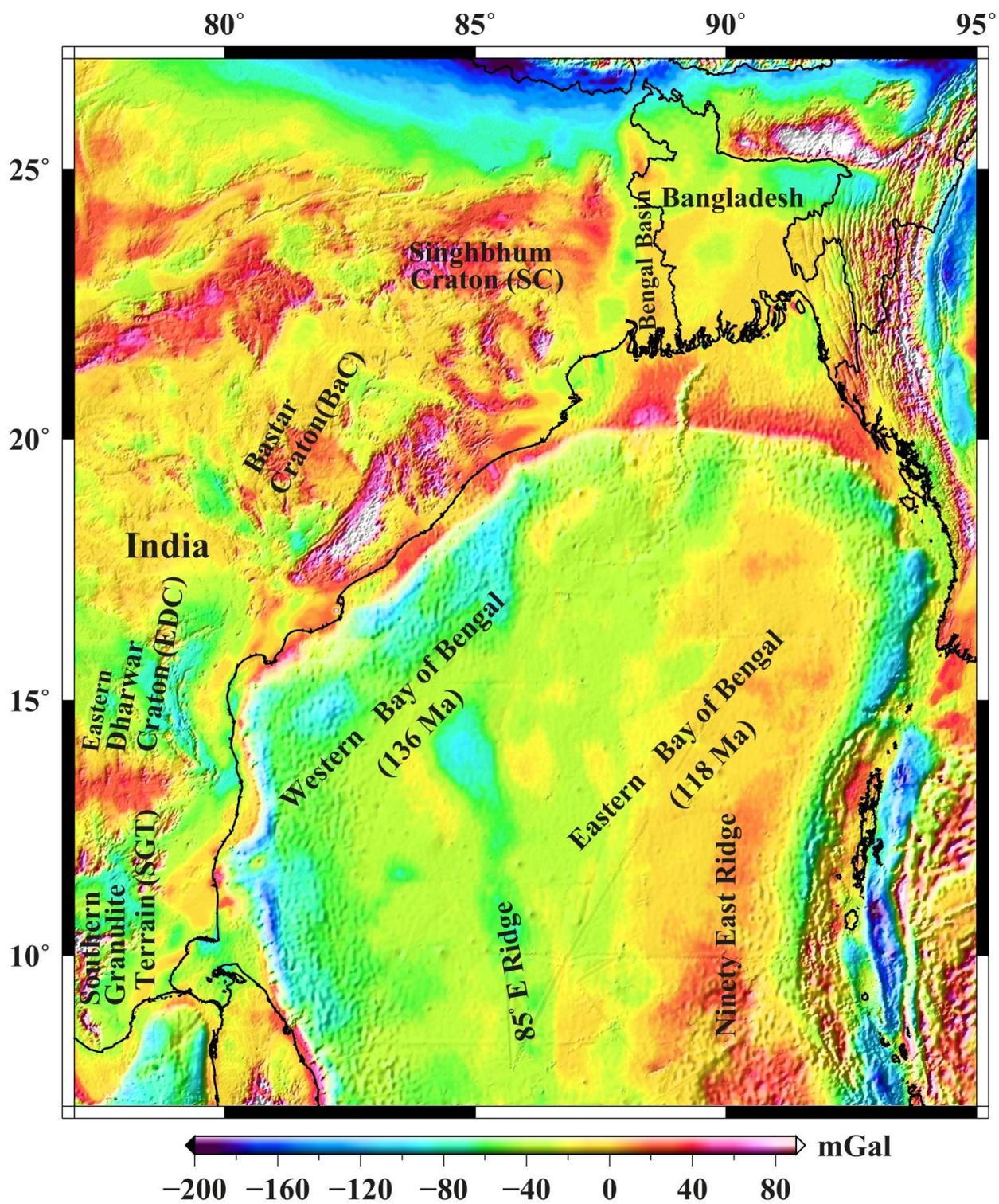

**Figure 1:** Geological map of the study region depicting individual terrains of the region superimposed on the free air gravity anomaly.

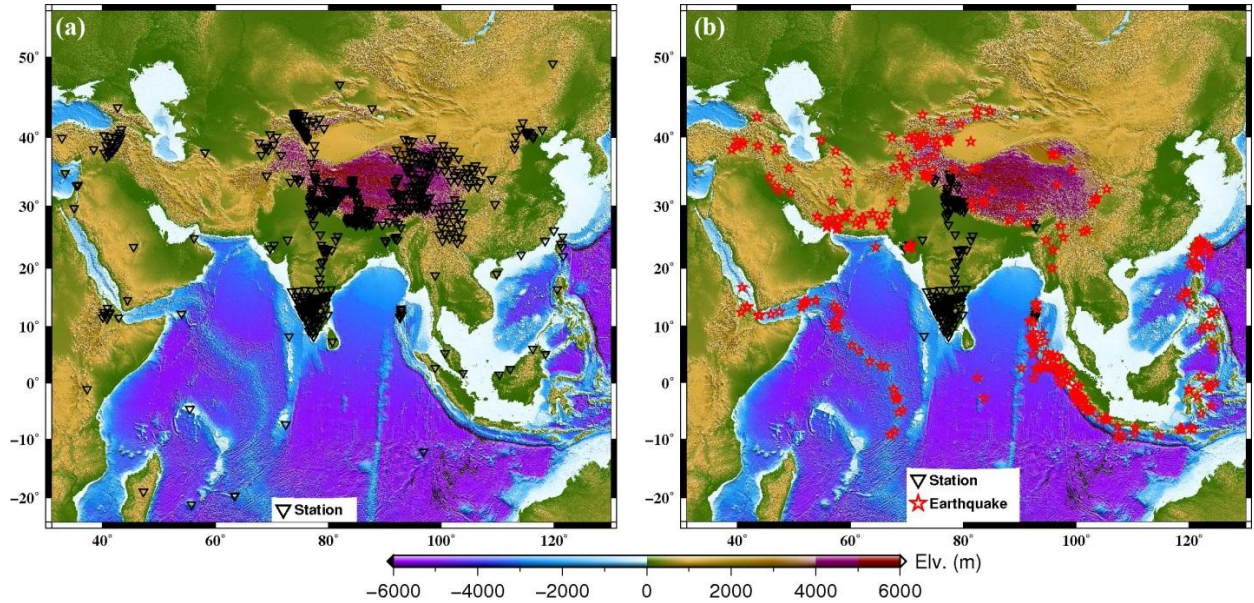

**Figure 2:** (a) Map shows the distribution of 683 seismic stations from India and the neighboring countries used to compute cross-correlation function from the vertical component ambient noise recording. This contributed to more than 21600 ray paths. (b) Additional group velocity data were generated using 417 earthquakes (red star) and the corresponding 209 seismic stations (black reverse triangles).

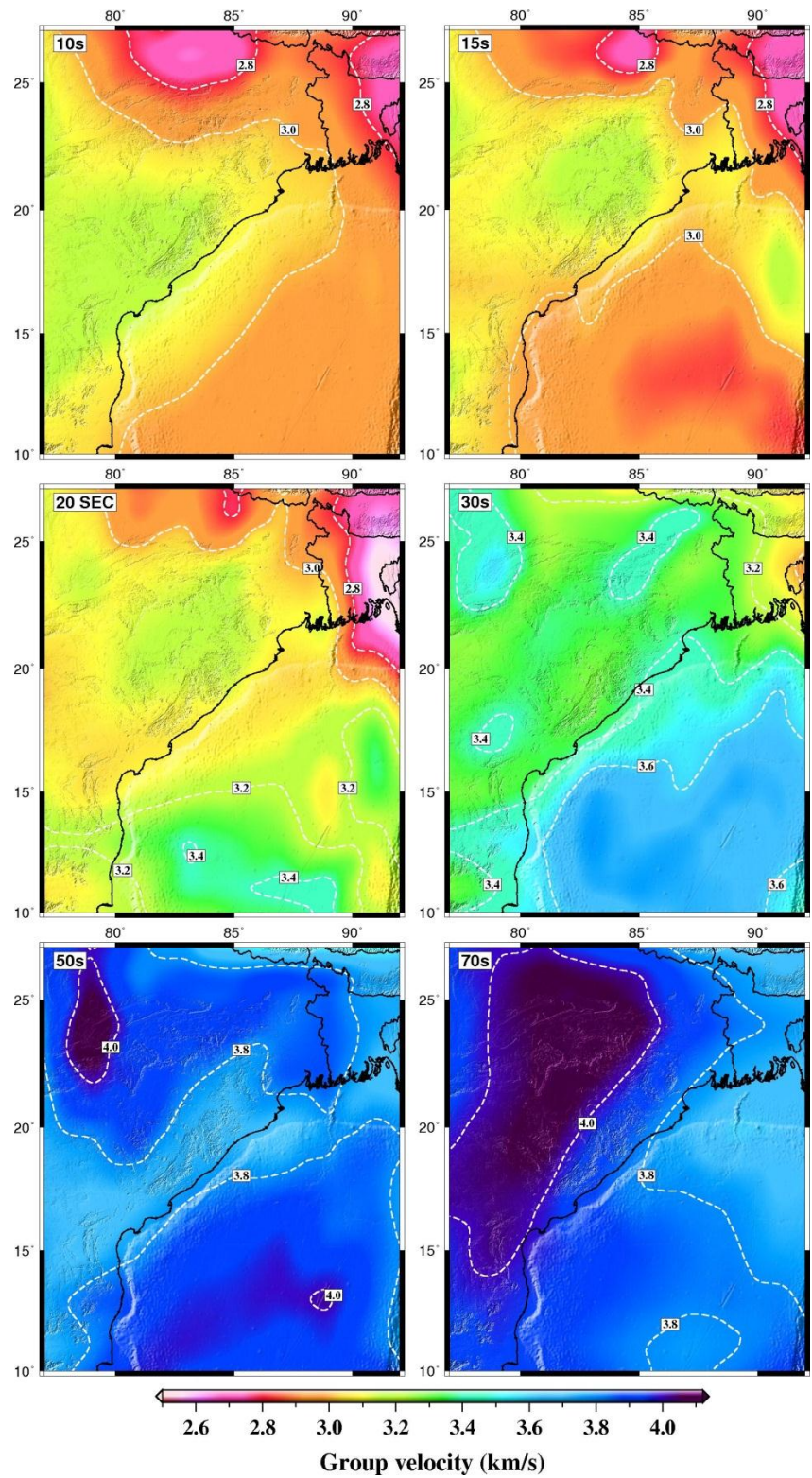

**Figure 3:** Fundamental mode Rayleigh wave group velocity maps at periods of the 10s, 15s, 20s, 30s, 50s, and 70s. Group velocity is contoured at an interval of 0.2 km/s.

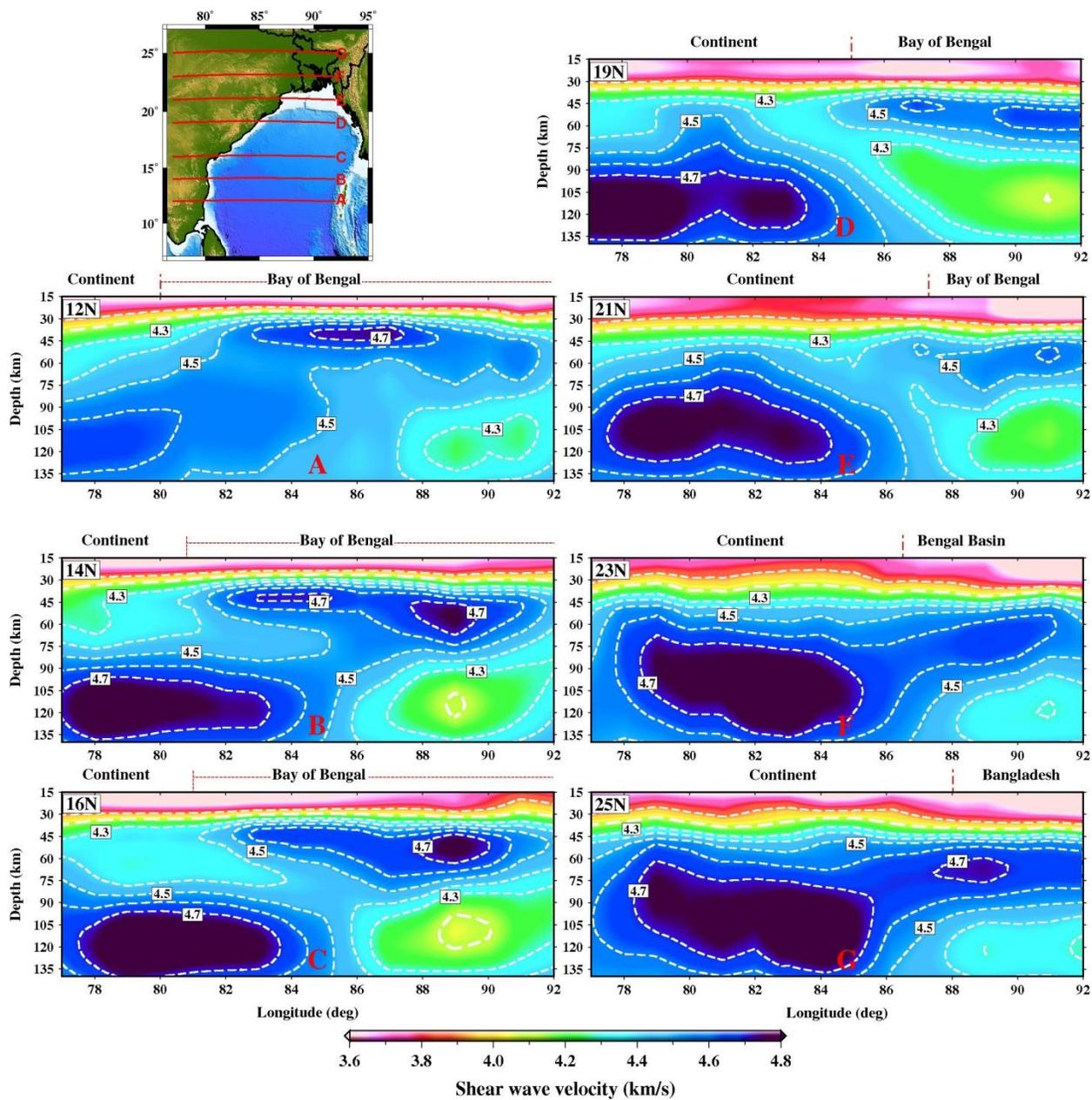

**Figure 4:** Shear wave velocity-depth section along west to east from Indian craton to the Bay of Bengal/Bangladesh. Velocity sections are generated at different latitudes from $12^{\circ}$ N to $25^{\circ}$ N (marked as profile from A to G). Shear wave velocity is contoured at an interval of 0.2 km/s. Major geological domains and tectonic boundaries are marked on top of individual figures.

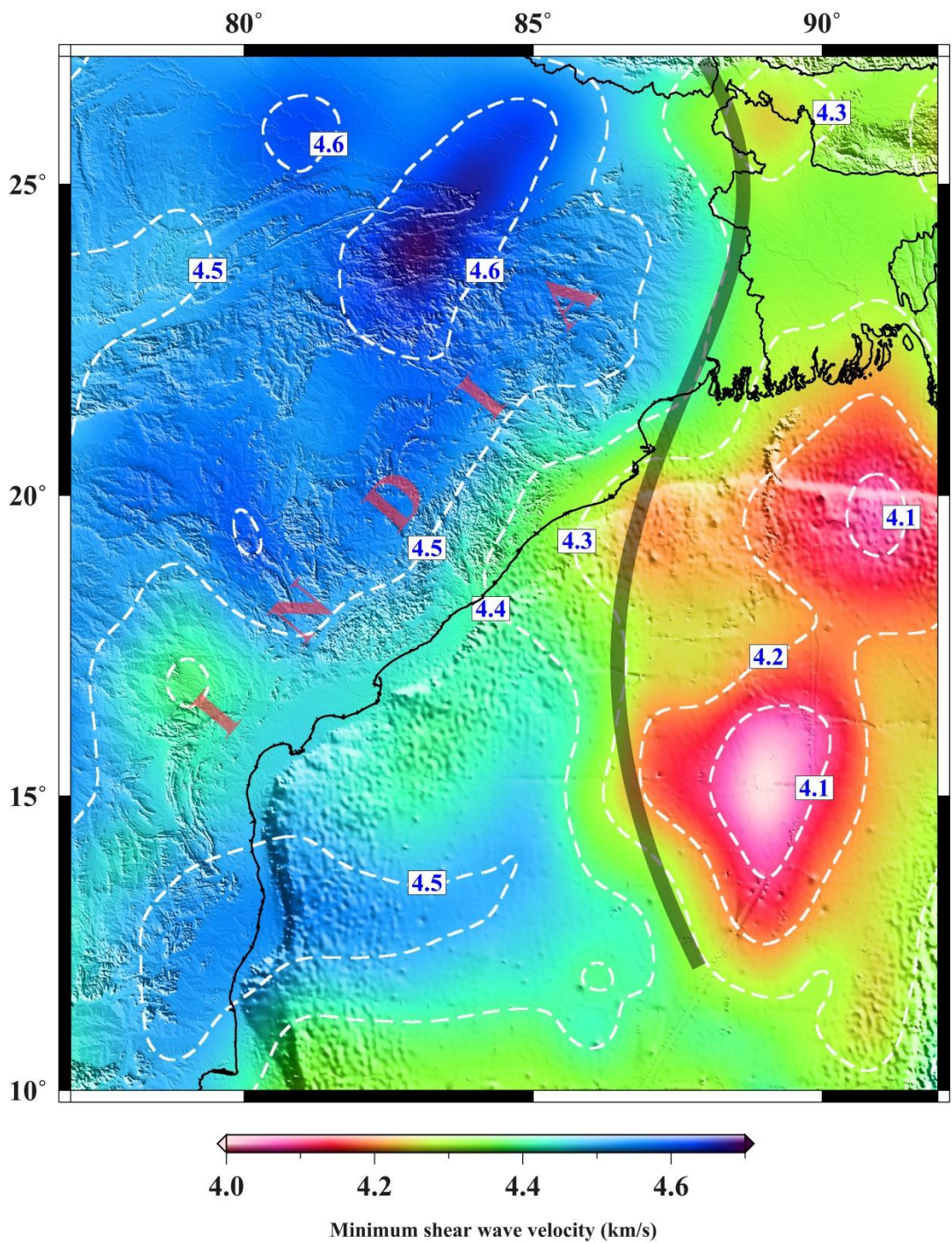

**Figure 5:** Map view of minimum shear wave velocity in the upper mantle beneath Bay of Bengal and adjoining Indian craton. Note the continuity of high velocity mantle beneath the Indian craton into the western Bay of Bengal. Interpreted Indian continent-ocean lithosphere boundary is shown by a thick black line.

# Supplemental Material

**Submerged Ancient Indian continent in the Bay of Bengal- inference from ambient noise and earthquake tomography**

Gokul Kumar Saha, S. S. Rai, K. S. Prakasa and V. K. Gaur

## Contents of this file

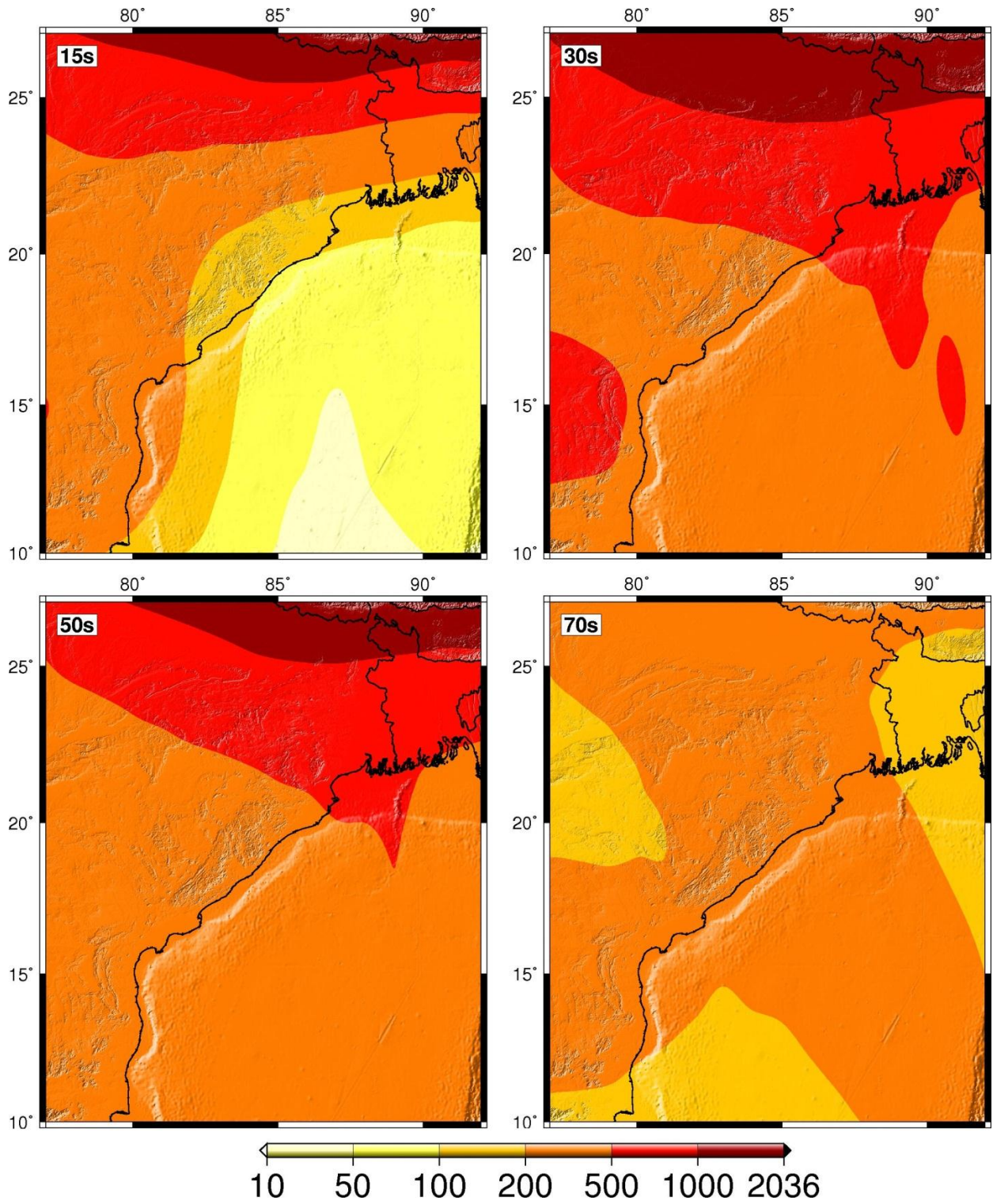

**Figure S1:** Ray path density distribution for combined noise and earthquake data computed for $1° \times 1°$ grid cells at periods of the 15s, 30s, 50s, and 70s.

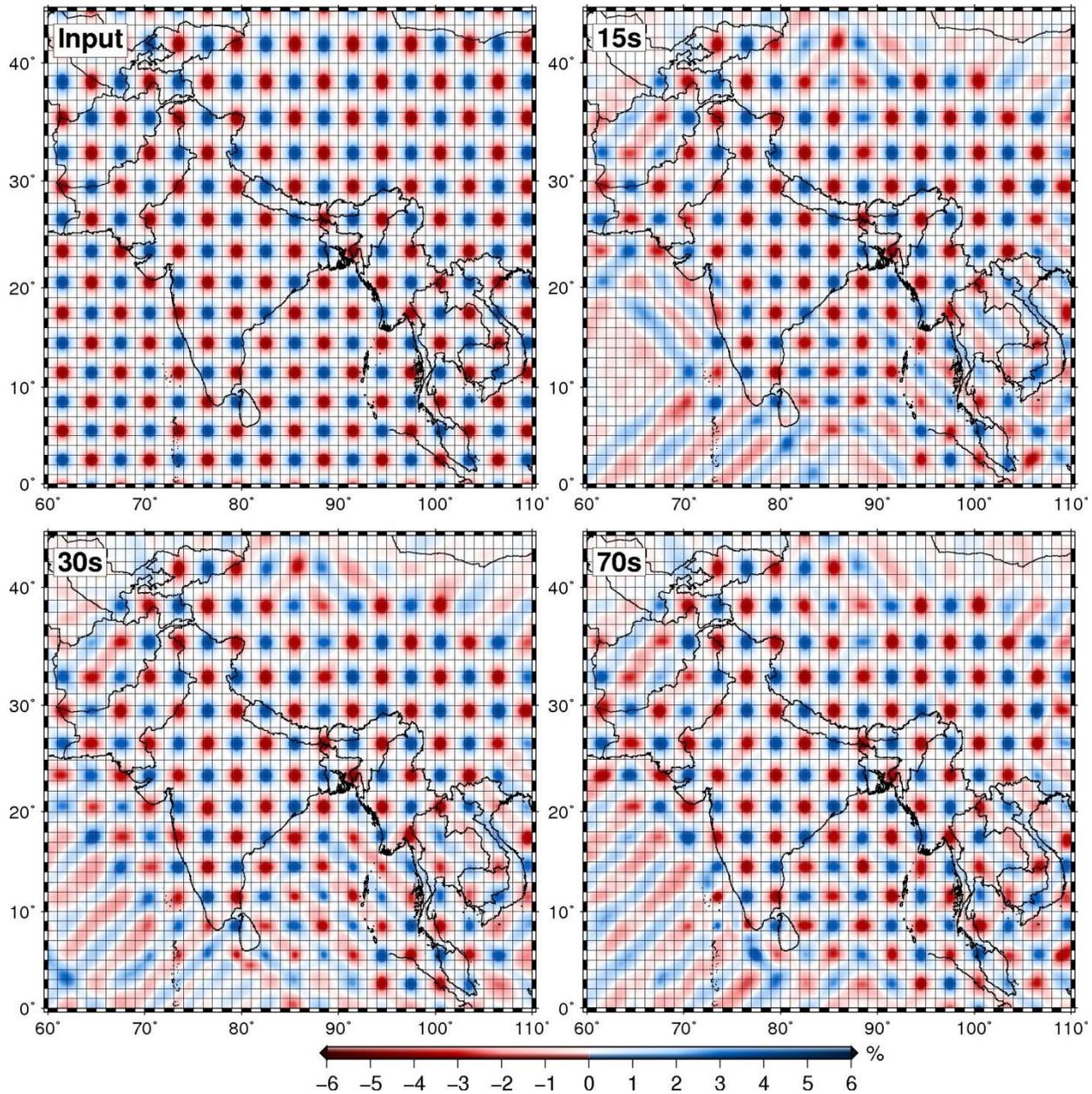

**Figure S2:** Checkerboard resolution tests for 15s, 30s and 70s Rayleigh wave group velocity maps with cells $1° \times 1°$. The input model is alternately $\pm6\%$ velocity variations in the blocks. Corresponding recovery maps are produced by inverting synthetic travel times through the checkerboard model for the same paths as those used in the tomographic inversion of real data.

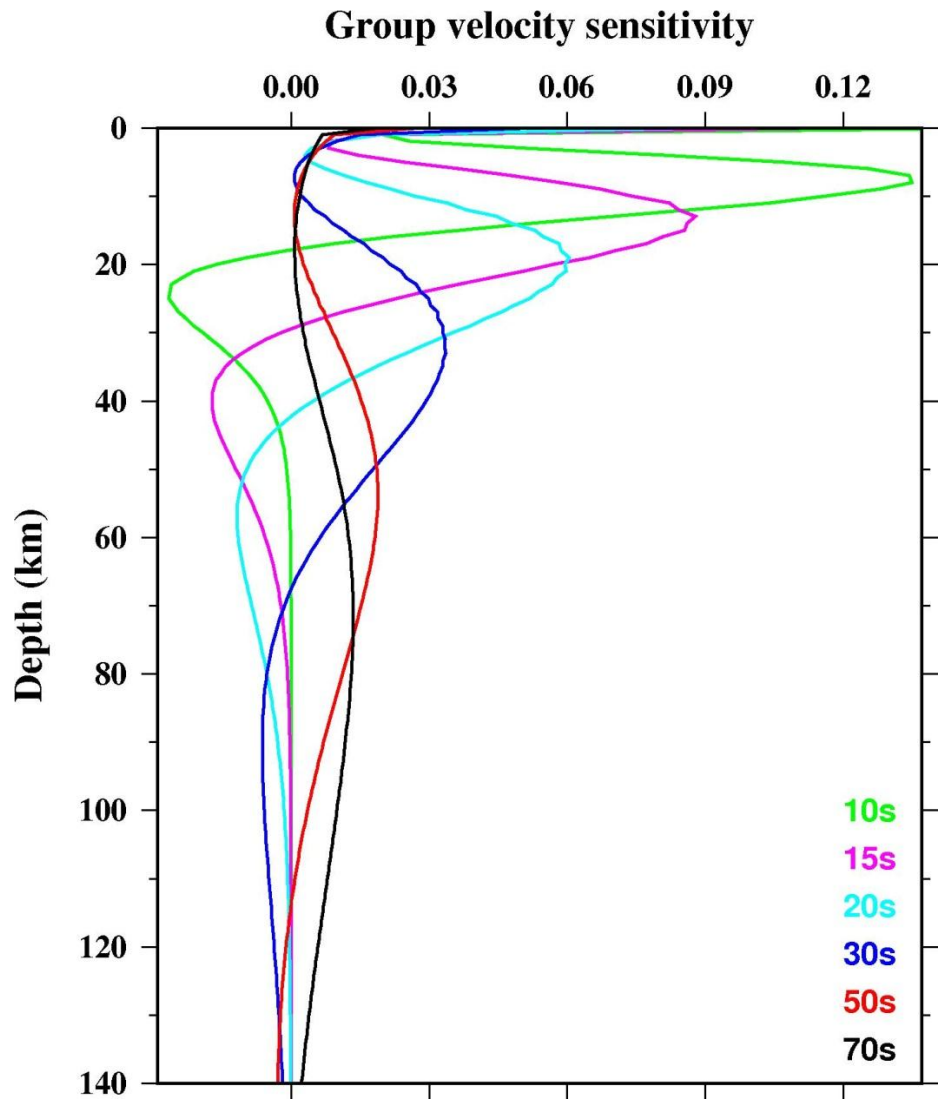

**Figure S3:** The depth sensitivity kernel of fundamental mode Rayleigh wave group velocities at time periods 10s 15s, 20s, 30s, 50s and 70s.

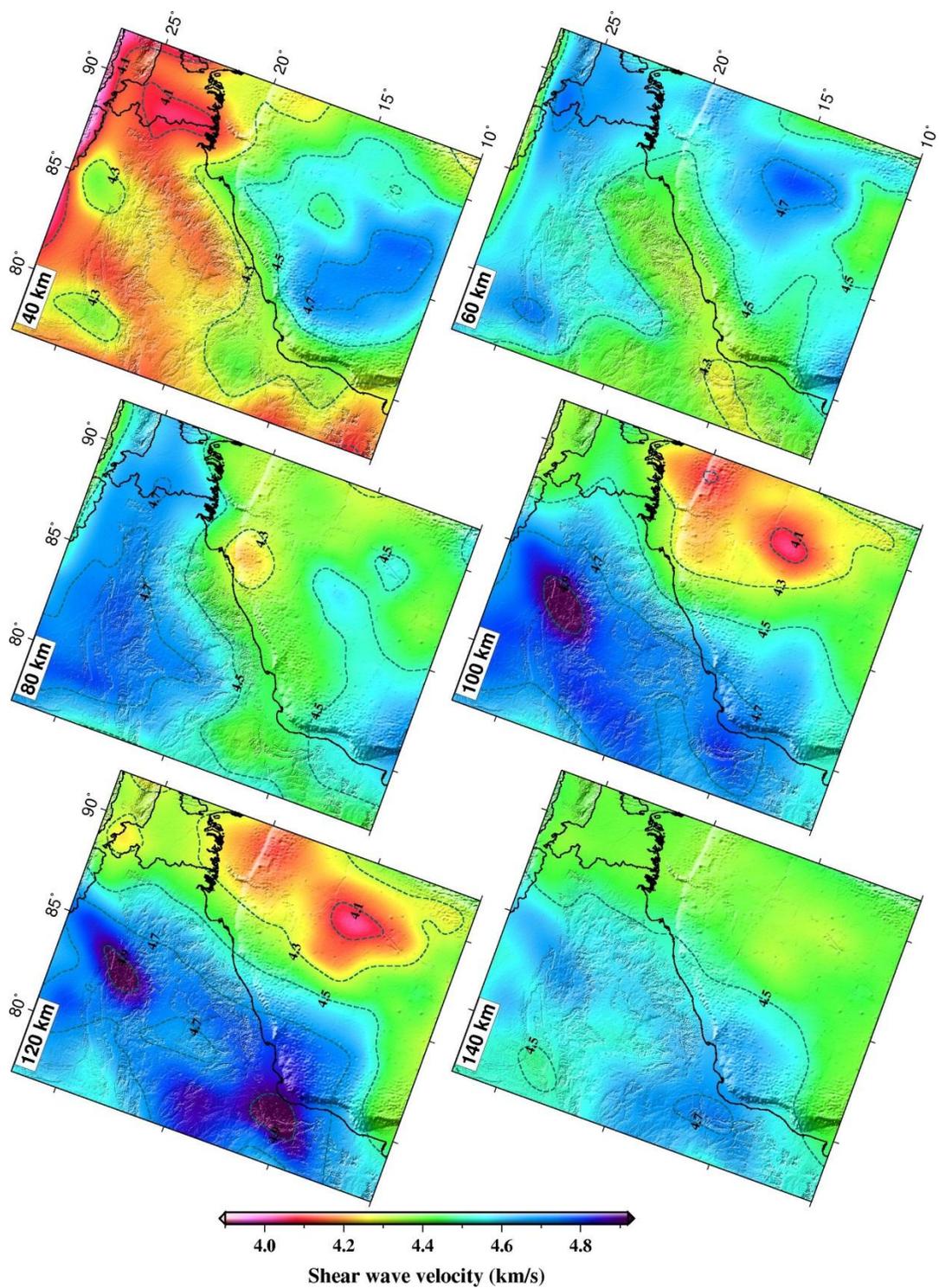

**Figure S4:** Map view of the shear wave velocity of the 3-D model at different depth. The depth for each map is marked in the upper-left corner of each panel. Shear wave velocity is contoured at an interval of 0.2 km/s.